\documentclass{moriond}

\def\Journal#1#2#3#4{{#1} {\bf #2}, #3 (#4)}

\def\NCA{\em Nuovo Cimento}

\def\NPB{{\em Nucl. Phys.} B}
\def\PLB{{\em Phys. Lett.}  B}
\def\PRL{\em Phys. Rev. Lett.}
\def\PRD{{\em Phys. Rev.} D}

\def\EPJC{{\em Eur. Phys. J.} C}
\def\EPL{{\em Europhys. Lett.}}
\def\ZETF{{\em Zh. Eksp. Teor. Fiz.}}

\def\be{\begin{equation}}
\def\ee{\end{equation}}
\def\bea{\begin{eqnarray}}
\def\eea{\end{eqnarray}}

\newcommand{\Photo}{\includegraphics[height=35mm]{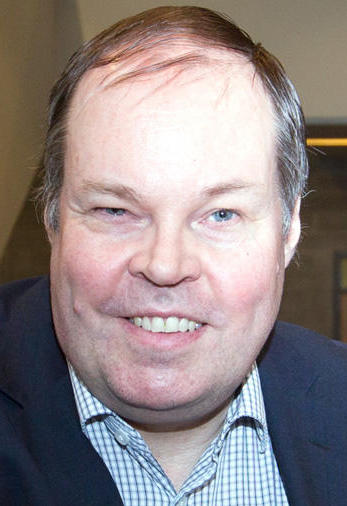}}

\begin{document}
\vspace*{4cm}
\title{COMPARISON OF PP AND P$\overline{\rm \bf P}$ DIFFERENTIAL ELASTIC CROSS SECTIONS AND OBSERVATION OF COLOURLESS C-ODD GLUONIC EXCHANGES}

\author{K. \"{O}STERBERG on behalf of the D0 and TOTEM collaborations}

\address{Helsinki Institute of Physics and Department of Physics, P.O. Box 64,\\ 
FI-00014 University of Helsinki, Finland}

\maketitle\abstracts{The TOTEM 2.76, 7, 8, and 13 TeV pp elastic differential cross sections are extrapolated using a data-driven approach to obtain the 1.96 TeV pp elastic cross section. A difference with a 3.4$\sigma$ significance is observed between the extrapolated pp elastic cross section and the D0 p$\bar{\rm p}$ elastic cross section at 1.96 TeV in the region of the diffractive minimum and the second maximum of the pp cross section, providing evidence for colourless C-odd gluonic exchanges, also denoted as odderon. These results are combined  with a TOTEM analysis of the same $t$-channel C-odd exchanges based on forward pp elastic scattering at TeV scale. The combined significance is larger than 5$\sigma$ and is interpreted as the first observation of odderon exchange.}

\section{Introduction}

High energy hadronic elastic scattering is traditionally described only by t-channel colourless multi-gluon (two or more) exchanges with an even charge parity (C), also denoted as the “pomeron”. However colourless C-odd multi-gluon (three or more) exchanges, “odderon”, are allowed~\cite{Lukaszuk} and even predicted by Quantum Chromodynamics~\cite{Braun}, but suppressed with respect to the dominating C-even exchanges. The amplitudes of C-even exchanges are predominantly imaginary, whereas C-odd exchanges are expected to mainly contribute to the real part of the amplitude. Contrary to the C-even amplitude, the C-odd amplitude has a different sign for proton-proton (pp) and proton-antiproton (p$\bar{\rm p}$). Hence at TeV scale where gluonic exchanges are dominant and meson exchanges negligible, any significant difference in the elastic differential cross section between pp and p$\bar{\rm p}$ at the same energy would be evidence for C-odd exchanges. C-odd exchanges are expected to be visible in the $\rho$ parameter, the ratio of real to imaginary part of the hadronic elastic amplitude at $|t|$ = 0, or where the imaginary part of the hadronic amplitude is suppressed e.g. at the diffractive minimum. Colourless multi-gluon exchanges at low $|t|$ values correspond to the exchange of (non-perturbative) colourless gluonic compounds rather than exchanges of colourless combinations of independent gluons coupling to different partons in the proton. The odderon has been searched for during the last 50 years, however convincing experimental evidence has up to now been missing.
 
TOTEM has measured the elastic pp cross section at 2.76~\cite{totem276TeV}, 7~\cite{totem7TeV}, 8~\cite{totem8TeV} and 13 TeV~\cite{totem13TeV} by measuring both intact protons at CERNs Large Hadron Collider (LHC) using Roman Pots (RPs) equipped with silicon detectors placed 210-220 m from the interaction point. D0 has measured the elastic p$\bar{\rm p}$ cross section at 1.96 TeV~\cite{d0196TeV} at Fermilab's Tevatron collider using similar techniques with RPs equipped with scintillating fibre detectors situated 23-31 m from the interaction point. 

\begin{figure}
\begin{minipage}{0.51\linewidth}
\centerline{\includegraphics[width=0.995\linewidth]{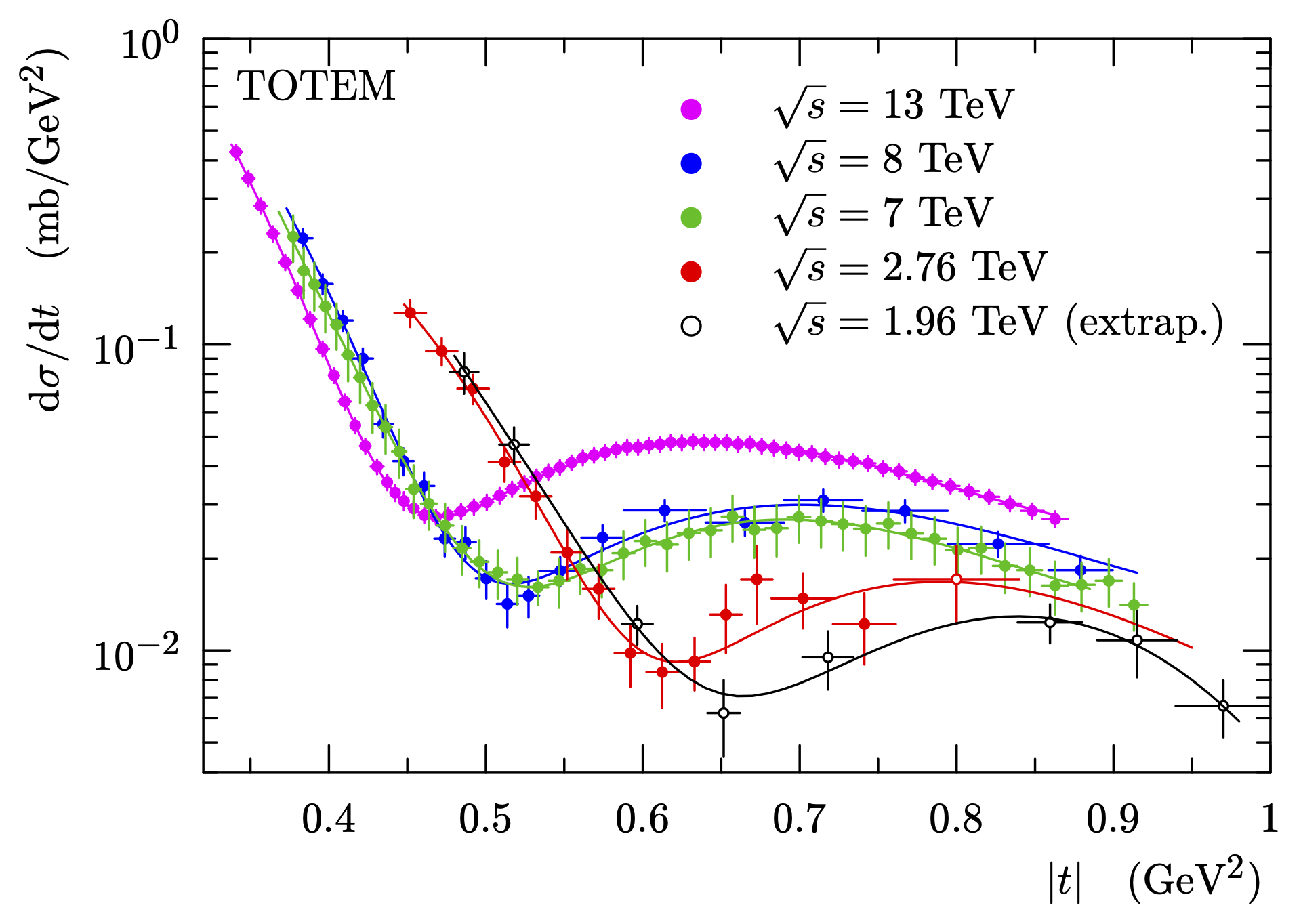}}
\end{minipage}
\hfill
\begin{minipage}{0.55\linewidth}
\vspace*{-0.5cm}
\hspace*{-0.35cm}
\centerline{\includegraphics[width=0.995\linewidth]{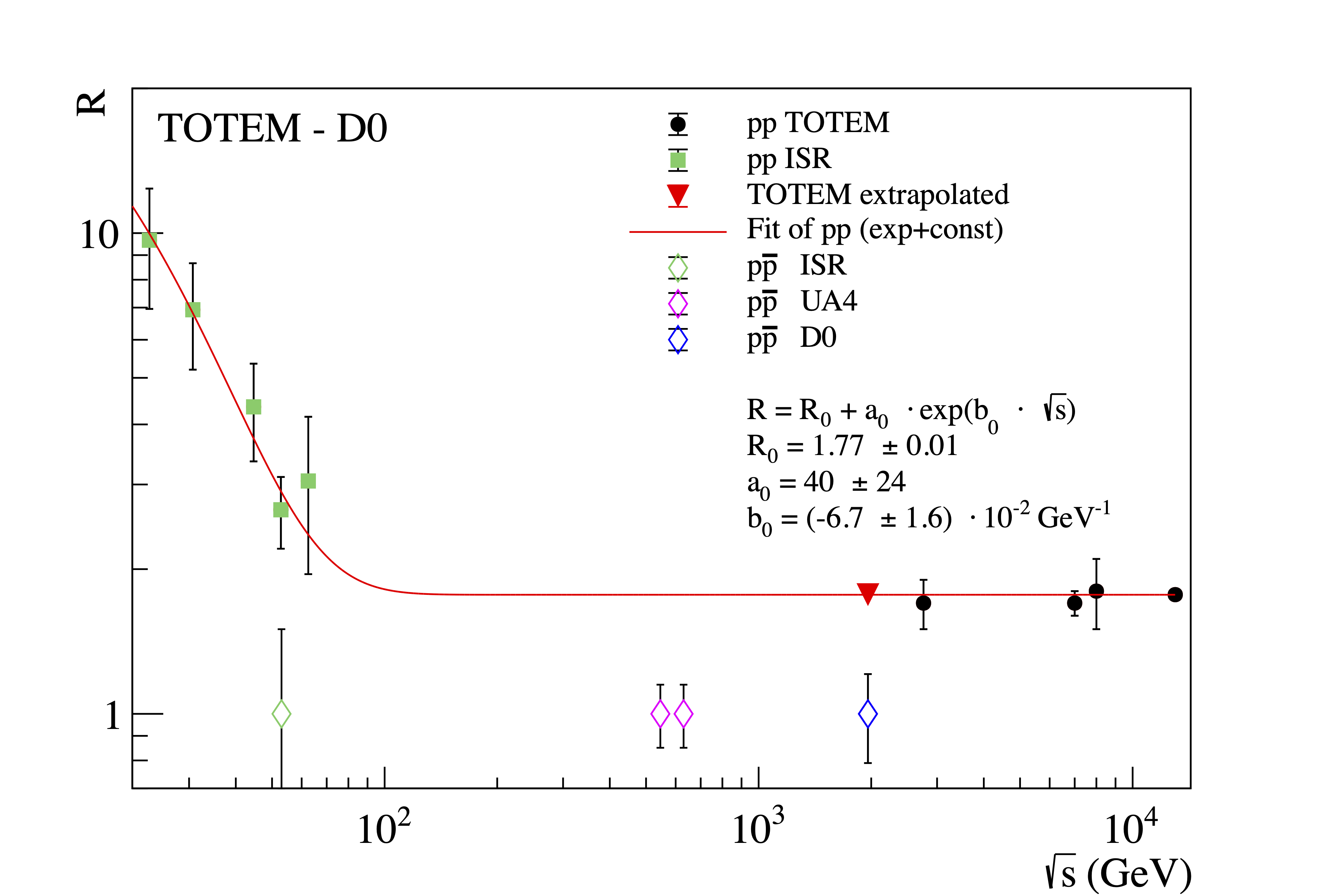}}
\end{minipage}
\hfill
\caption[]{Left: The TOTEM pp elastic cross sections at 2.76, 7, 8, and 13 TeV (full circles), and
the extrapolation to 1.96 TeV (empty circles). Right: The ratio, R, of the cross sections at the bump and
dip for pp and p$\bar{\rm p}$ data.}
\label{fig:figure1}
\end{figure}

The pp elastic differential cross sections at TeV scale show characteristic diffractive minima (``dips''), followed by secondary maxima (``bumps''), as shown by Fig.~\ref{fig:figure1} (left), whereas the p$\bar{\rm p}$ elastic differential cross sections only exhibits ``kinks'' at the expected position of the dips. Fig.~\ref{fig:figure1} (right) shows the ratio R between the cross section at the bump and the dip for pp and p$\bar{\rm p}$ elastic data. A descreasing R of pp as a function of $\sqrt{s}$ can be observed below 100 GeV that flattens out at LHC energies. Since the p$\bar{\rm p}$ elastic data contain no visible dip and bump, the R-values are estimated from the maximal ratio over neighbouring bins centered on the expected bump and dip locations predicted by the pp data. Assuming that the flat behaviour of R for the pp cross sections continues down below 2 TeV, the R-values of pp and p$\bar{\rm p}$ differ by more than 3$\sigma$.

\section{Extrapolation of pp data to 1.96 TeV and comparison of pp and p$\bar{\rm \bf p}$ data}

\begin{figure}
\begin{minipage}{0.43\linewidth}
\centerline{\includegraphics[width=0.975\linewidth]{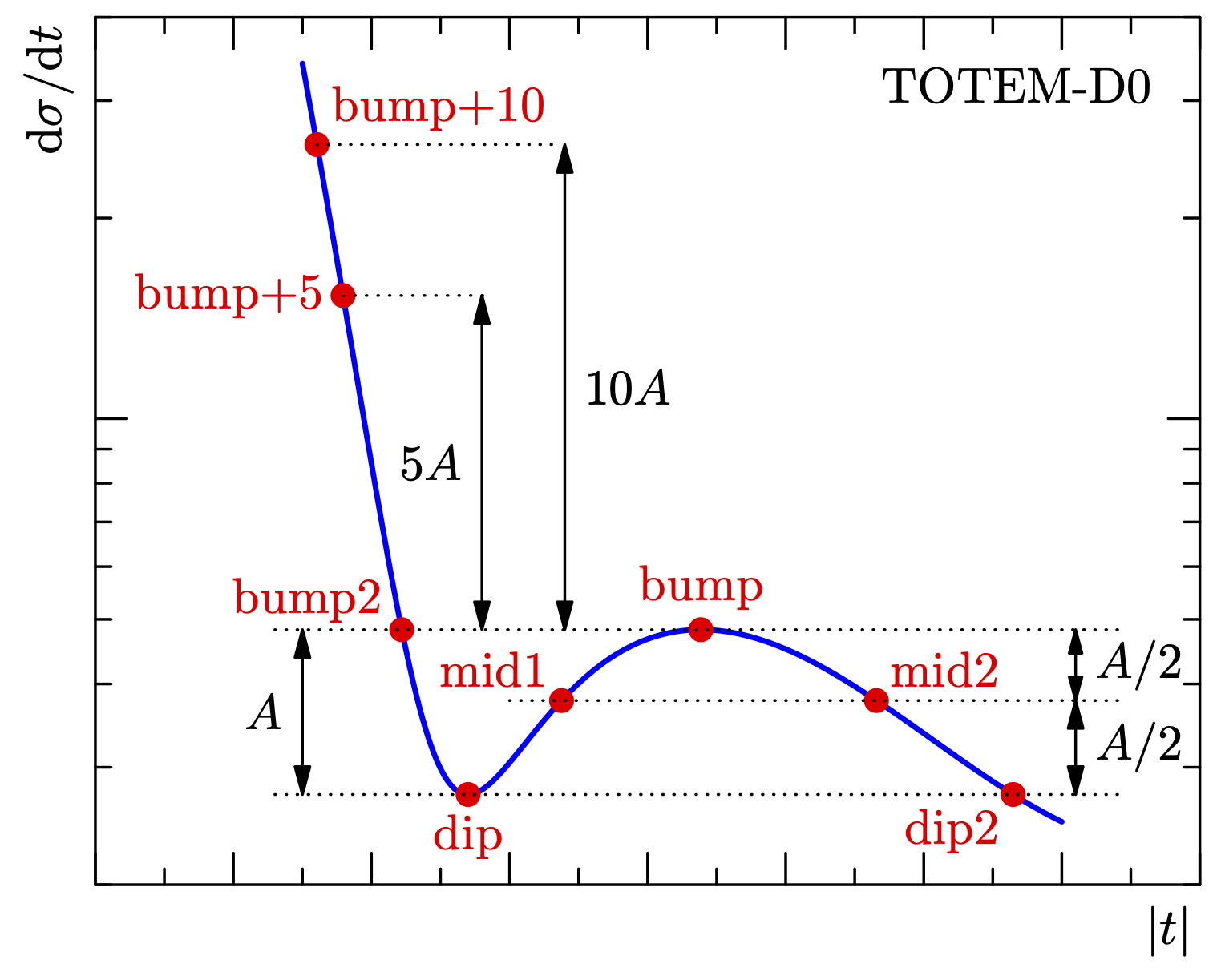}}
\end{minipage}
\hfill
\begin{minipage}{0.57\linewidth}
\centerline{\includegraphics[width=0.975\linewidth]{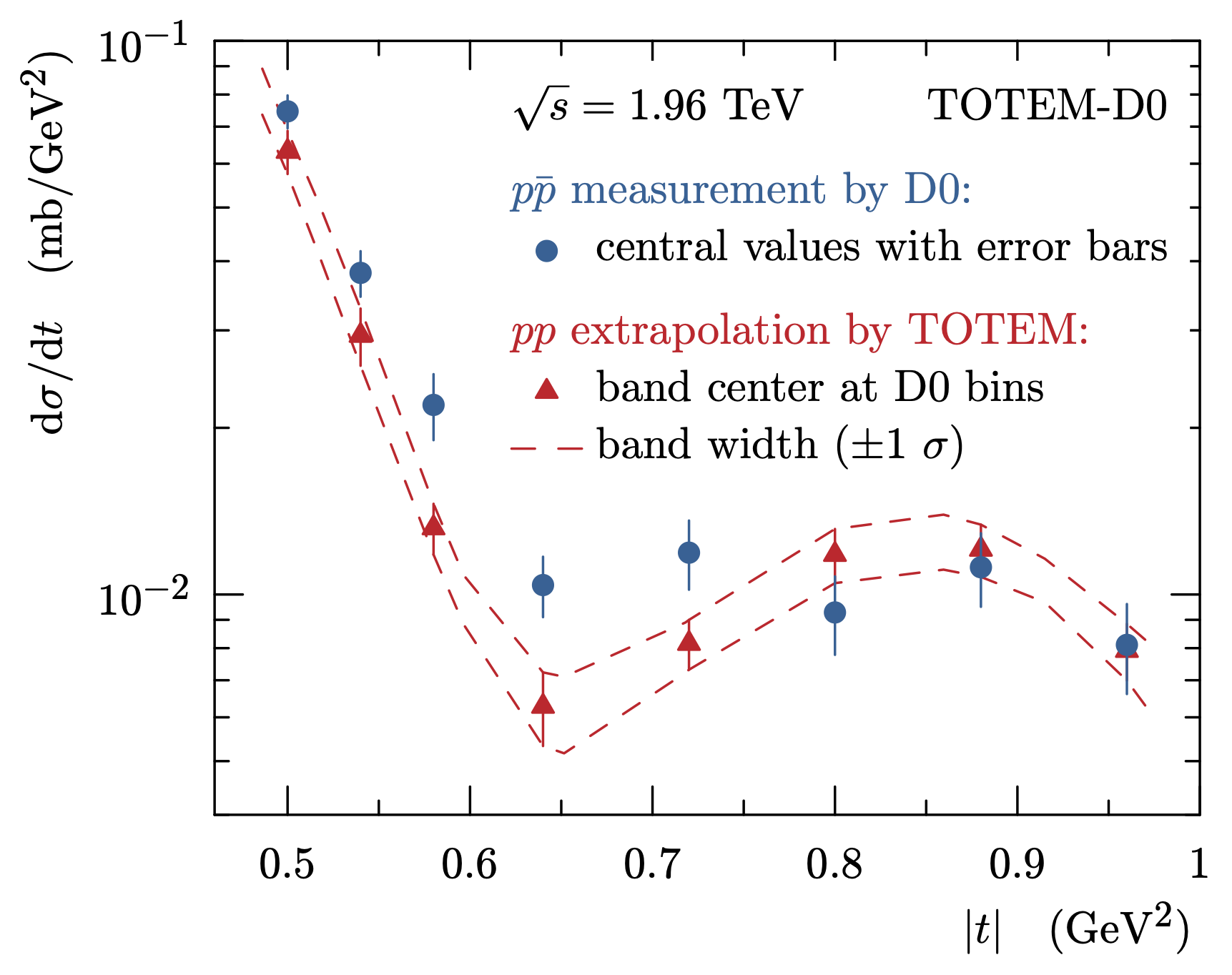}}
\end{minipage}
\hfill
\caption[]{Left: Definition of the characteristic points in the TOTEM differential cross section data. Right: Comparison between the D0 p$\bar{\rm p}$ measurement at 1.96 TeV and the extrapolated TOTEM pp cross section, rescaled to match the D0 OP. The dashed lines show the 1$\sigma$ uncertainty band on the extrapolated pp cross section.}
\label{fig:figure2}
\end{figure}

To obtain a more quantitative estimate of the pp and p$\bar{\rm p}$ difference, eight characteristic points of the pp elastic cross section as shown by Fig.~\ref{fig:figure2} (left) are selected and both their $|t|$ and $d\sigma/dt$ values are extrapolated to 1.96 TeV using their values of the characteristic points of the TOTEM measurements at 2.76, 7, 8 and 13 TeV. The values of $|t|$ and $d\sigma/dt$ as functions of $\sqrt{s}$ for each characteristic point are fitted using $|t| = a \, log(\sqrt{s} [{\rm TeV}]) + b$ and $(d\sigma/dt) = c \, \sqrt{s}[{\rm TeV}] + d$ respectively. The parameter values are determined for each characteristic point separately and the same functional form describes the dependence for all characteristic points. The reduced $\chi^2$ values of most fits are close to one. The above forms are chosen for simplicity after it has been checked that alternative forms providing adequate fits yielded similar extrapolated values within uncertainties. The $|t|$ and $d\sigma/dt$ values of the characteristic points extrapolated to 1.96 TeV are shown as open black circles in Fig.~\ref{fig:figure1} (left).

To estimate the 1.96 TeV pp elastic $d\sigma/dt$ at the same $|t|$ values as the D0 p$\bar{\rm p}$ data, the extrapolated pp data are fitted with a double exponential, where the first exponential describes the cross section up to the dip and the second exponential the cross section at the bump and the subsequent decrease. The double exponential also provides a good fit for all measured TOTEM pp cross sections as shown by the fitted functions in Fig.~\ref{fig:figure1} (left). The uncertainty of the extrapolated pp cross section at the $|t|$ values of the D0 p$\bar{\rm p}$  data are estimated using Monte Carlo (MC) simulations in which the cross section values of the eight characteristic points are varied within
their Gaussian uncertainties and new double exponential fits are performed. 

Finally, the extrapolated pp cross section is scaled so that at the optical point (OP), the elastic $d\sigma/dt$ at $|t|$ = 0, is the same as that for p$\bar{\rm p}$. The OP cross sections at high energies are expected to be equal if there are only C-even exchanges~\cite{sigmatot}. In addition, the logarithmic slopes of the elastic diffractive cone are expected to be the same~\cite{slope}. As this is the case within uncertainty for the pp and p$\bar{\rm p}$, the scaling is constrained to preserve the measured logarithmic slopes. To obtain the pp OP at 1.96 TeV, the TOTEM pp total cross section measurements ($\sigma_{tot}$) at 2.76, 7, 8, and 13 TeV are extrapolated to 1.96 TeV using $\sigma_{tot}$ = $b_1 \, log^2(\sqrt{s} [{\rm TeV}]) + b_2$ that is only valid above 1 TeV. This gives a $\sigma_{tot} = 82.7 \pm 3.1$ mb at 1.96 TeV. It has been checked that a polynomial formula or adding an additional $log \sqrt{s}$ term in the functional form leads to similar results well within uncertainties. The extrapolated total cross section is converted using the optical theorem to an OP cross section for pp of $357 \pm 26$ mb/GeV$^2$ using 
$\rho$ = 0.145 based on the COMPETE extrapolation~\cite{compete}. The systematic uncertainty due to possible C-odd exchange effects on the OP of 2.9 \% is estimated from maximal odderon scenarios~\cite{maxodd} and added in quadrature to the experimental uncertainty. A single exponential fit of the D0 measured p$\bar{\rm p}$ $d\sigma/dt$ gives an OP cross section of $341 \pm 49$ mb/GeV$^2$ that is in good agreement with the obtained pp OP cross section. The comparison between the final extrapolated and rescaled TOTEM pp cross section at 1.96 TeV and the D0 p$\bar{\rm p}$ measurement is shown in Fig.~\ref{fig:figure2} (right).

A $\chi^2$ test is made comparing the measured  p$\bar{\rm p}$ data to the rescaled pp data shown in Fig.~\ref{fig:figure2} (right), normalized to the integral cross section of the  p$\bar{\rm p}$ measurement in the examined $|t|$-range of 0.50 to 0.96 GeV$^2$. The fully correlated OP normalization and logarithmic slope of the elastic cross section are added as separate terms to the $\chi^2$ sum. The correlations for the D0 measurements at different $|t|$-values are small, but the correlations between the eight TOTEM $d\sigma/dt$ at the $|t|$ values of the D0 data are large due to the double exponential fit used to obtain them. Taking into account the constraints on the normalization and logarithmic slopes, the $\chi^2$  test with six degrees of freedom gives a significance of 3.4$\sigma$. A cross check of the result giving a compatible significance is made using a modified Kolmogorov-Smirnov test, where  the covariance matrices are included by MC methods and the difference of the integrated cross section in the examined $|t|$-range is added via the Stouffer method~\cite{stouffer}.

The observed difference in the pp and p$\bar{\rm p}$ elastic cross section is interpreted as colourless C-odd gluonic exchanges. In
agreement with predictions~\cite{maxodd}$^{,}\, $\cite{durham}, the pp cross section shows a deeper dip and stays below the p$\bar{\rm p}$ cross section until the bump region as seen from Fig.~\ref{fig:figure2} (right).

\section{Combination with evidence from forward pp elastic scattering at multi-TeV scale}

The present result is combined with independent TOTEM evidence of C-odd exchanges based on the measurements of $\rho$ and $\sigma_{tot}$ for multi-TeV pp interactions. These variables are sensitive to differences in pp and p$\bar{\rm p}$ scattering that in turn in the multi-TeV range are related to the existence of C-odd gluonic exchanges. The TOTEM $\rho$ and $\sigma_{tot}$  results are incompatible with models having only C-even exchanges and provide independent evidence of C-odd exchange effects~\cite{rho13TeV} based on observations in completely different $|t|$ domains and TOTEM data sets.

The significances of the different measurements are combined using the Stouffer method~\cite{stouffer}. The $\chi^2$ for
the $\sigma_{tot}$ measurements at 2.76, 7, 8 and 13 TeV is computed with respect to the predictions given
from models without C-odd exchanges~\cite{compete}$^{,}\, $\cite{durham} including also model uncertainties when specified. Same was done separately for the TOTEM $\rho$ measurement at 13 TeV. Unlike the COMPETE models~\cite{compete}, the Durham model~\cite{durham} provides the predicted differential cross section without C-odd exchange contributions. Therefore a direct comparison of the predicted Durham model cross section at 1.96 TeV with D0 p$\bar{\rm p}$ data is used for the combined significance instead of the D0-TOTEM comparison. The 1.96 TeV cross section of the model is chosen since it is most sensitive to C-odd exchanges after the model has been tuned to the LHC elastic pp data.

The comparison of $\rho$ and $\sigma_{tot}$ with the set of models gave significances ranging between 3.4 and 4.6$\sigma$.  When these are combined with the above comparison of pp and p$\bar{\rm p}$ elastic cross sections, the significances range
from 5.2 to 5.7$\sigma$ for t-channel C-odd gluonic exchanges for all examined models~\cite{compete}$^{,}\, $\cite{durham}. In particular, for the model favored by COMPETE (RRP$_{nf}$L2$_u$)~\cite{compete}, the TOTEM $\rho$ measurement at 13 TeV provided a 4.6$\sigma$ significance, leading to a total significance of 5.7$\sigma$ for t-channel C-odd exchanges when combined with the above pp and p$\bar{\rm p}$ comparison.

\section{Summary}

In the present analysis~\cite{d0totem}, the D0 p$\bar{\rm p}$ elastic differential cross section at 1.96 TeV is compared to the TOTEM pp cross sections extrapolated to 1.96 TeV from measurements at 2.76, 7, 8, and 13 TeV using a data-driven approach. The two cross sections at 1.96 TeV differ with a significance of 3.4$\sigma$ providing already evidence that t-channel colourless C-odd gluonic exchanges, i.e. odderon, is needed to describe elastic scattering at high energies. When combined with an independent TOTEM C-odd exchange evidence based on $\rho$ and total cross section measurements in pp using completely different TOTEM data sets and $|t|$ domains, the significance is in the range 5.2 to 5.7$\sigma$ thus constituting the first experimental observation of C-odd gluonic exchanges or odderon.

\end{document}